# Global oscillations of the Sun according to the data of coronagraph SOHO LASCO C3

V.I. Efremov[a], L.D. Parfinenko[a], A.A. Soloviev[a,b] [*]

[a]Central (Pulkovo) astronomical observatory of Russian Academy of Science, Saint-Petersburg, 196140, Russia

[b]Kalmyk state university, Elista, 358000, Russia

*e-mail: solov@gao.spb.ru

**ABSTRACT**

We investigate the long-period fluctuations in the brightness of the Sun as a star using the measurements of sunlight reflected from the planets (Jupiter, Mars) when the light hits the field of view of the LASCO C3 coronagraph (Large Angle and Spectrometric Coronagraph Experiment).

## 1. INTRODUCTION

Low-frequency quasi-harmonic oscillations of the magnetic field of sunspots and associated magnetic elements, found both in terrestrial and cosmic observations, are one of the most important properties of these solar magnetic structures and represent a qualitatively new phenomenon in solar physics. Research of this kind of oscillation offers great opportunities for the diagnosis of the physical parameters of active formations on the Sun.

At present, there are a number of independent observations confirming the existence of long-period oscillations of a number of physical parameters of sunspots with periods of tens and hundreds of minutes. This is manifested in the variations of the magnetic field and the line-of-sight velocities of sunspots (Efremov et al. 2007; 2010; 2012), in the time-variations of the microwave emission of radio sources above the sunspots (Chorley et al., 2010); Smirnova et al., 2011). These oscillations probably reflect the quasi-periodic displacements of sunspots as an whole, well integrated magnetic structures (Solov'ev & Kirichek, 2014). Studies of long-period oscillations of the magnetic fields of sunspots were carried out (Efremov et al. 2010, 2012, 2014) based on the data of space instruments MDI (SOHO) and HMI (SDO). These studies have shown that the limiting low-frequency mode of global oscillations of the magnetic field of a sunspot as a whole is apparently the mode with a period in the 800-1800 minute band (13-30 hours). Its period depends essentially and in a nonlinear manner on the magnitude of the magnetic field of the sunspot. In addition to this extremely low mode, the higher modes with periods in the bands 40-45, 60-80, 135-170, 220-240 and 480-520 minutes were also revealed in the power spectra of sunspots, and the oscillation power in these bands monotonously and rapidly decreases with increase in frequency, which is characteristic of overtones arising as a result of the nonlinear nature of the oscillations. The limiting oscillatory mode stably exists in sunspots for many days, but at the same time during every 1.5-2 days its power increases. This time coincides with the average lifetime of the super-granule cell.

Using the continuous 5-day time series of filtergrams obtained in the Hα line on the global network of GONG telescopes, the ultra low-frequency modes with a period of 20-30 hours have also been detected for the long-lived dark filaments lying on the solar disk (Efremov et al., 2016). Foullon et al. (2009), using the data of SOHO/Extreme-Ultraviolet Imaging Telescope (EIT) in line 195 Å, have discovered the ultra low-frequency modes of brightness oscillations of coronal loops, with a period of 10-30 hours.

The question arises, is there a connection between these low-frequency oscillations of active magnetic structures observed in the solar atmosphere and solar gravitational *g*-modes? In contrast to the well-studied acoustic *p*-modes, the solar gravitational modes provide information on deep solar layers, on the structure and dynamics of the solar core. Until now, no reliable observations of *g*-modes have been presented, it is believed that they are very closely concentrated to the solar core and cannot be fixed on the surface. But it is interesting that the theory gives an estimates from 16 minutes up to 28 hours (Appourchaux et al., 2010) for a possible interval of periods of gravitational modes. This interval falls within the range of periods of low-frequency oscillations of solar activity elements discovered by us and other researchers. Perhaps these g-modes, due to resonant phenomena, contribute to the buildup of low-frequency oscillations of active magnetic structures on the surface of the Sun. It is also possible that these modes could manifest themselves as global oscillations of the Sun as a star. Hence, there is interest in the study of time-variations in the brightness of sunlight reflected from the surface of the planets.

Recently P. Gaulme et al. (Http://arxiv.org/abs/1612.04287v1) and Jason F. Rowe et al. (Http://arxiv.org/abs/1702.02943v1), using the data of Kepler K2 space telescope, made a careful monitoring of the brightness of reflected sunlight from the planet Neptune. In the power spectra of reflected sunlight, they found three periods of approximately 4, 8 and 13 hours, to which they attributed the solar origin.

In this paper, we use the data of Lasco C3 cosmic coronagraph (Large Angle and Spectrometric Coronagraph Experiment) (https://lasco-www.nrl.navy.mil/). The continuous 5-day time series of brightness of two planets have been treated.

**2. METHODS AND OBSERVATIONS**

For studying the time variations of the solar radiation reflected from the planets, to search the periodic components in the sunlight, we used both the traditional Fourier expansion and the methods developed for the study of non-stationary time series and nonlinear systems. These are methods that use the procedure of singular decomposition SVD (Singular Value Decomposition). Among these, are used such methods as the Multi-Taper Method (MTM), the Proper Orthogonal Decomposition (POD), Hilbert-Huang transform (HHT, EMD algorithm), CaterPillarSSA (Golyandina et al., 2001). Of these methods, we prefer the latter because of a number of advantages. The most important of which is the choice of the lag "manually", i.e. the number of eigen functions of the expansion is given by the researcher himself. This moment is the key in the study of time series with weak separability of additive components, i.e. when the eigenvalues of the trajectory matrix of decomposition are very close. This makes the CaterPillarSSA method more attractive than the others. In the program implementation of this method, the main components of the decomposition can be investigated and, what is especially important in the further reconstruction of the original series, they can be visualized and ordered by increasing their contribution to the original series. The phase diagrams directly show the nature of the circulation process of the components and make it possible to build up the time series, excluding one or the other component.

Time series of the flux change of sunlight from the planets were formed using the method of extreme value developed by us earlier (Efremov et al., 2010). The essence of it is as follows: the object on the FITS image of the Lasco C3 coronagraph is immersed into a strip from which it does not exit during the observation. Next, the procedure is parallelized (when two processes run in parallel): if the object is a distributed one and is subject to a changes during the observation, the binarization procedure is performed above a certain threshold (Nobuyuki Otsu, 1979), and the area of the object is calculated as the number of beans above the specified threshold, and its center is defined. The value of the flux is taken from the central pixel. If the object is a point or close to it, then the extreme value of the flux inside the strip is simply calculated, i.e. the flux in a point is taken.

## 3. RESULTS

As an observational material for detecting periodic changes in the solar radiation reflected from the planets, the sequences of filtergramms with the cadence of 12 minutes were taken from the data archive of SOHO/LASCO space observatory, with a size of 1024 × 1024 pixels. For the initial study, a limited number of outer planets were selected: Mars, Jupiter. The duration of the observation session, as a rule, was 5 days. The trend components for this observation interval were deleted from the studied time series and not considered.

**3.1 Mars 2015/05/27-31**

The general course of the change in the flux of sunlight reflected from Mars in the period from May 27 to May 31, 2015, expressed in arbitrary flow units of the SOHO/LASCO instrument, is shown in Fig.1.

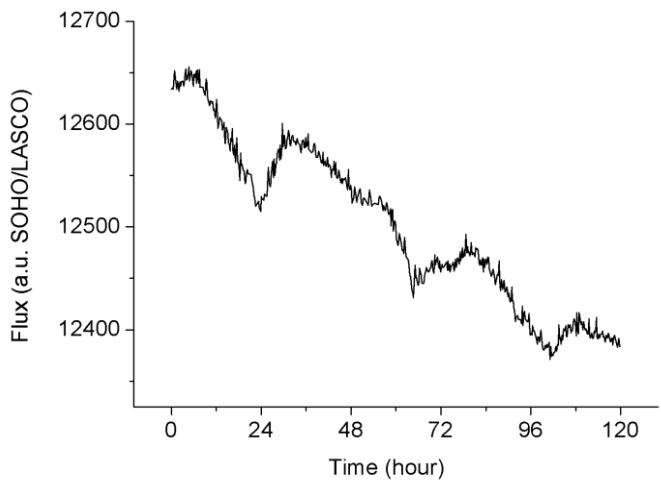

.

Figure 1. The original time series of variations of the sunlight flux reflected from the planet Mars.

After removing the trend component, the decomposition of the original time series, performed by the CaterPillarSSA program, shows that only 3 phase diagrams of the main components of the decomposition arranged in descending amplitudes correspond to quasi-periodic modes present in the studied time series. The periods of these 3 modes are: 36-38 hours (M36h), 24-26 hours M24h) and 8-10 hours (M8h). Figure 2 shows the detrended time series itself and its three main components with periods. The possible cause of the appearance of these periodic components in the reflected light of the planet will be discussed below, but for the time being we emphasize the stability of the amplitudes of these modes during the observation time.

Considering shorter realizations (daily intervals of observations), we found that in the high-frequency part of the spectrum there is another very stable period with a value of 80 minutes. It belongs to a special class of artifacts (p2p) associated with the moving of a continuously distributed object through a discrete receiver matrix (Efremov et al., 2010). It is a technical artifact and it is not caused by any physical process associated with solar phenomena. In the present observations, this period is far in the high-frequency part of the spectrum of fluctuations of the sunlight flux, and further we do not consider it.

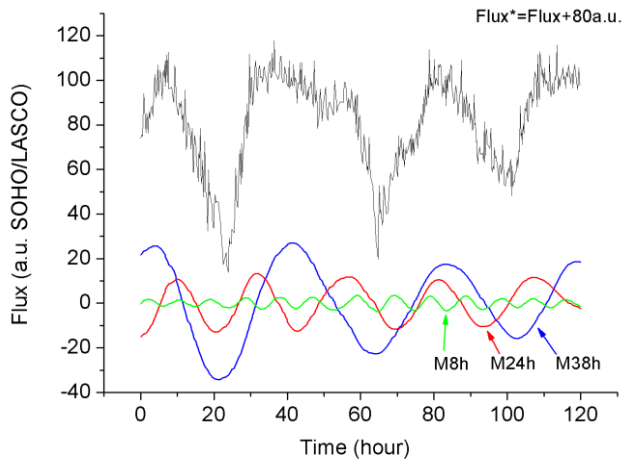

Figure 2. Detrended time series of variations of the sunlight flux reflected from the planet Mars and its three main decomposition components: modes with periods of 36-38 hours, 24-26 hours and 8-10 hours.

## 3.2. Jupiter  2016/09/17-21

The observation interval was 106 hours. Phase diagrams of the main components of the decomposition of the time series of the change in the sunlight flux reflected from Jupiter show that only three quasi-periodic components are present in the original time series. Fig. 3 shows the initial time series after the removal of the trend and its three main components of the decomposition with periods of 36-38 hours, 15 hours and 8-10 hours, respectively.

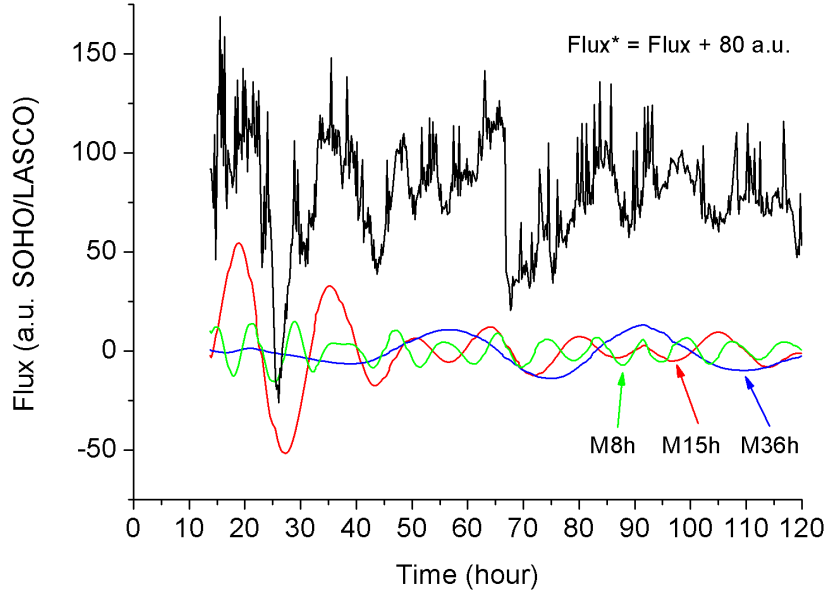

Figure 3. Detrended time series of variations of the sunlight flux reflected from the planet Jupiter and its three main decomposition components: modes with periods: 36-38 hours, 15 hours and 8-10 hours.

## 4. DISCUSSION

We see that some quasi-periodic components in the time series of the variations of sunlight reflected from the planets Mars and Jupiter are identical: two modes with periods of 36-38 hours and 8-10 hours were surprisingly repeated. Let us give a possible explanation of the physical cause of the periods obtained. For the planet Mars, the appearance of a period of modulation of reflected sunlight at 24-26 hours is logical to explain the manifestation of surface photometric inhomogeneities of the planet. The fact is that the diurnal period of its rotation is 24.2 hours and may well become the cause of the appearance of a corresponding circulation component. As for the other 2 modes: 36-38 hours and 8-10 hours, they, apparently, are of solar origin. The same two modes were repeated in the observation of Jupiter. We note that the ratios of amplitudes of M38h and M8h modes are practically preserved: as before, the amplitude falls to shorter periods, which is typical for the power spectra of hollow resonators. However, it should be noted that there is some difficulty in explaining the M8h mode for Jupiter, since the daily period of revolution of Jupiter is 9 hours and 40 minutes, which falls within the specified range of the periods of the mode, and its appearance in the time series of variations may be due to the diurnal rotation of the planet as to be of solar origin. For a harmonic with a period of 15 hours, we did not find a reasonable explanation, at least we can assume that this mode is not of solar origin, because there is no such component for the planet Mars.

Thus, apparently, only the M36h mode with a period of 36-38 hours has a solar origin without controversy. Note that among the manifestations of various activities on the surface of the Sun, there is a process with a similar period - this is a phenomenon of supergranulation.

**CONCLUSION**

5-day continuous temporal series of brightness of the planets of Mars and Jupiter were studied. They found matching modes: 8-10 hours and 36-38 hours, which, apparently, are of solar origin. In addition, 25 day-long series of temporal variations of reflected solar radiation were obtained for Mars. In them, an even more low-frequency 250-hour mode was discovered. The amplitudes of the detected modes decrease from low frequencies to high frequencies.